\documentclass[twocolumn,fleqn,10pt]{wlscirep_e}
             
\title{Control of tunneling in a triple-well atomtronic switching device}

\author[1]{Karin Wittmann Wilsmann}
\author[2]{Leandro H. Ymai}
\author[2]{Arlei Prestes Tonel}
\author[3,*]{Jon Links}
\author[1]{Angela Foerster}
\affil[1]{Instituto de F\'isica da UFRGS, Av. Bento Gon\c{c}alves 9500, Agronomia, Porto Alegre, RS, Brazil}
\affil[2]{Universidade Federal do Pampa, Av. Maria Anuncia\c{c}\~ao Gomes de Godoy 1650, Bairro Malafaia, Bag\'e, RS, Brazil}
\affil[3]{School of Mathematics and Physics, The University of Queensland, Brisbane, QLD 4072, Australia.}

\affil[*]{jrl@maths.uq.edu.au}
\usepackage{epsfig}
\usepackage{amsmath}
\usepackage{amsfonts}
\usepackage{graphicx}
\usepackage{subfigure} 
\usepackage{amssymb}
\usepackage{graphics}
\usepackage{psfrag}
\usepackage{ulem}

\title{Control of tunneling in an atomtronic switching device \\}

\begin{abstract}
\noindent {\bf 
The precise control of quantum systems will play a major role in the realization of atomtronic devices. As in the case of electronic systems, a desirable property is the ability to implement switching. 
Here we show how to implement switching in a model of dipolar bosons confined to three coupled wells. The model  describes interactions between  bosons, tunneling of bosons between adjacent wells, and the effect of an external field. We conduct a study of the quantum dynamics of the system to probe the conditions under which switching behavior can occur. The analysis considers both integrable and non-integrable regimes within the model. Through variation of the external field, we demonstrate how the system can be controlled between various ``switched-on'' and ``switched-off'' configurations.} 
\end{abstract}

\begin{document}
\flushbottom
\maketitle
\thispagestyle{empty}

The phenomenon of quantum tunneling is paramount in many studies of ultracold quantum gases. The two-well Bose-Hubbard Hamiltonian has been very successful in modeling quantum tunneling \cite{milburn,l01}, displaying two principal dynamical scenarios. These are referred to as Josephson tunneling and self-trapping, and they have been experimentally observed \cite{albiez05}.  In the case of tunneling, the system can also be controlled to produce either alternating or direct currents \cite{llss07}. A three-well system opens up wider possibilities for physical behaviors \cite{cbzs11,cbcs12,busch}, most notably as an ultracold version of a transistor \cite{anderson_transistor}, or similar type of switching device. The individual wells can be identified as the source, gate, and drain, potentially forming a building block in the emerging field of atomtronics \cite{anderson_atomtronic,ob15,ab16}. This prospect is driving research into transistor-like structures beyond the electronic domain \cite{Olshanii,Zinner_transistor}. 
 
Here we investigate the influence of integrability in the control of tunneling in a triple-well system. 
To do so, we must go beyond the familiar three-well Bose-Hubbard model \cite{nhmm00,bfp09,gcw14,o14,ggmmj15}, and consider a more general system which facilitates an integrable limit. 
Such a model has already been introduced into the literature. It models  dipole-dipole interactions 
and the tunneling between adjacent sites for a population of ultracold dipolar bosons with large dipole moment, such as 
chromium or dysprosium, loaded in an aligned triple-well potential. The Hamiltonian has the general structure \cite{lps10}

\begin{align}
{\mathcal H}& = \frac{U_0}{2}\sum_{i=1}^{3}  N_i (N_i-1) + \sum_{i=1}^{3} \sum_{j=1; j\neq i}^{3} \frac{U_{ij}}{2} N_i N_j \nonumber \\ 
&\qquad -J_1(a_1^\dagger a_2 + a_1 a_2^\dagger) - J_3(a_2^\dagger a_3 + a_2 a_3^\dagger).
\label{hintless}
\end{align}
The canonical creation and annihilation operators, $a_i^\dagger$ and $a_i$, $i=1,2,3$, represent the three bosonic degrees of freedom in the model, and $N_i= a_i^\dagger a_i$, $i=1,2,3$ is the number operator for each well. The parameters $J_i,\, i=1,3$ are the couplings for the tunneling between wells, and  $U_0$ is the coupling constant for on-site interactions which results from contact interactions and dipole-dipole interactions (DDI). Both of these can be either attractive or repulsive, which in principle allows for the manufacture of weak net on-site interaction. The parameters  $U_{ij}=U_{ji}, \,i\neq j$ characterize DDI between particles on different sites. Although the DDI follows an inverse cubic law, it is also dependent on the angle between dipole orientation and the displacement between dipoles. In combination with the geometry of the trap potential (viz. oblate versus prolate), it is entirely feasible to adjust the system parameters across a wide range of values. Importantly, this includes the possibility for the inter-well couplings $U_{ij}$ to have greater magnitude than the on-site coupling $U_0$. The experimental feasibility of this system for dipolar bosons was detailed by Lahaye et al.\cite{lps10},  using a triple-well potential.  The wells are aligned along the $y$-axis, separated by a distance $l$, with bosons polarized by a sufficiently large external field along the $z$-direction. It was shown that $U_{12}=U_{23}=\alpha U_{13}$, where the parameter $4\leq \alpha \leq 8$ depends only on the ratio $l/\sigma_x$, where $\sigma_x$ is the width of the Gaussian cloud along the $x$-direction. (See Methods for further details.)

In the case when $U_{13}=U_0$, the Hamiltonian (\ref{hintless}) is integrable\cite{ytfl17}. In this limit there exists an additional conserved operator besides the Hamiltonian and the total particle number, such that the number of independent conserved operators is equal to the number of degrees of freedom.  While for classical systems integrability is well-known to prohibit chaotic behavior,  the consequences for quantum system are less understood \cite{cm11,l13}. Notwithstanding, it is recognised that quantum integrability has far reaching impacts. 
One route to characterize the degree of chaoticity in a quantum system is through energy level spacing distributions \cite{lyar16}. Integrable systems tend to display Poissonian distributions \cite{bt77}, while non-integrable systems generally observe the Wigner surmise \cite{w} following the Gaussian Orthogonal Ensemble, or similar  \cite{d62,pzbmm93}. 
Another impact of quantum integrability is the absence of thermalisation, observed in a quantum version of Newton's cradle \cite{weiss} and similar systems \cite{lea}. Here we will demonstrate how integrability, and the breaking of it, can be utilised to investigate tunneling dynamics. This work contrasts the above mentioned studies in that it applies to a system with very low number of degrees of freedom.  

A simple means of breaking integrability in the model is through an applied external field. Generally, it might be expected that this will drive the system into a chaotic dynamical regime. However it is shown that in certain circumstances the changing dynamics of the model, through tuning of the external field, can be predicted with remarkable accuracy. 
The result can be understood by revealing the structure of a hidden subsystem within the model. 
This level of control points towards the potential utility of a physical realisation of the model as a quantum switch. 

\section*{Results}
\subsection*{Integrability}
It has been established that the model (\ref{hintless}) contains a family of integrable multi-well tunneling models when $U_{13}=U_0$ \cite{ytfl17}. In this case, we can write the Hamiltonian in the reduced form \break 
$\displaystyle H_0=-{\mathcal H}+{(\alpha+1)U_0}N^2/4-{U_0}N/2$ yielding 
\begin{align}
H_0&=U(N_1-N_2+N_3)^2 \nonumber \\ 
&\qquad + J_1(a_1^\dagger a_2 + a_1 a_2^\dagger) + J_3(a_2^\dagger a_3 + a_2 a_3^\dagger),
\label{hint}
\end{align}
where $U=(\alpha-1)U_0/4$.
Note that (\ref{hint}) commutes with the total number operator
$N=N_1+N_2+N_3$,
and the interchange of the indices
$1$ and $3$ leaves the Hamiltonian invariant.
The Hamiltonian has, beyond the energy and the total number of particles $N$, another independent conserved quantity  expressed through the operator \cite{ytfl17}
\begin{equation}
Q = J_1^2 N_3 + J_3^2 N_1 - J_1 J_3(a_1^\dagger a_3+a_3^\dagger a_1). \label{charge}
\end{equation}
This conserved operator can alternatively be interpreted as a tunneling Hamiltonian for a two-well subsystem containing only wells 1 and 3. Because $Q$ admits the factorization $Q=\Omega^\dagger \Omega$, where $\Omega = J_1 a_3 - J_3 a_1$, the dynamical evolution governed by $Q$ is harmonic for any initial state. Later, it will be shown that $Q$ assumes a fundamental role in the analysis of resonant \cite{peter} quantum dynamics of the system (\ref{hint}).  This arises due to an unexpected connection with virtual processes. Details are provided in Methods.

As the model has three degrees of freedom and three independent conserved quantities satisfying 
\begin{eqnarray}
[H_0,N]=0, \;\;\;\;\; [H_0,Q]=0,  \;\;\;\;\; [N,Q]=0,
\end{eqnarray}
the model is integrable. Further details about the integrability, and associated exact solvability, have been established. This was achieved through the Yang-Baxter equation and associated Bethe Ansatz methods \cite{ytfl17}.

\subsection*{Breaking the integrability}

\begin{figure}[h!]
\centering
\includegraphics[width=6cm]{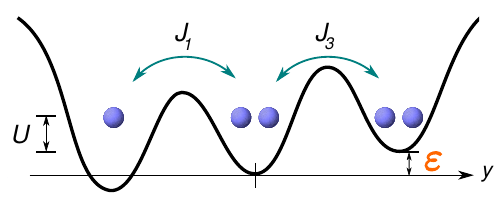}
\caption{Schematic representation of the system. With reference to the Hamiltonian $H=H_0+H_1$, the arrows $J_1$ and $J_3$ represent the tunneling couplings between the wells, $U$ characterizes inter-well and intra-well interaction between bosons, while $\epsilon $ is the coupling strength for an external field.}
\label{esquema}
\end{figure}

In order to break the integrability we add to the Hamiltonian (\ref{hint})  the operator $H_1= \epsilon (N_3-N_1)$,
which acts as an external field for the wells labeled 1 and 3. 
This is schematically shown in Fig. \ref{esquema}. It is important to observe that the above Hamiltonian still commutes with
the operator $N$. However, the operator $Q$ is not conserved because the commutator
$[H,Q]= 2\epsilon J_1 J_3 (a_1^\dagger a_3 - a_3^\dagger a_1)$
is non-zero when the parameters $\epsilon $, $J_1$ and $J_3$ are all non-zero.

\subsection*{Structure of energy levels}
The integrable three-well system (\ref{hint}) possesses many features in common with the two-well Bose-Hubbard model, which is also integrable because the total number operator is conserved and there are only two degrees of freedom. Set $J=\sqrt{J_1^2+J_3^2}$. Following Leggett\cite{l01}, it is useful to define the regimes:
\begin{itemize}
\item[] {\it Rabi}:  $U \ll JN^{-1}$. 
\item[] {\it Josephson}: $JN^{-1} \ll U \ll JN$.
\item[] {\it Fock}: $JN \ll U$. 
\end{itemize}
where in the two-well case the ``Fock regime corresponds to a
strongly quantum pendulum, while in the Rabi and Josephson
regimes the behavior is (semi)classical''\cite{l01}.

\begin{figure}[h!]
\centering
\includegraphics[width=8cm]{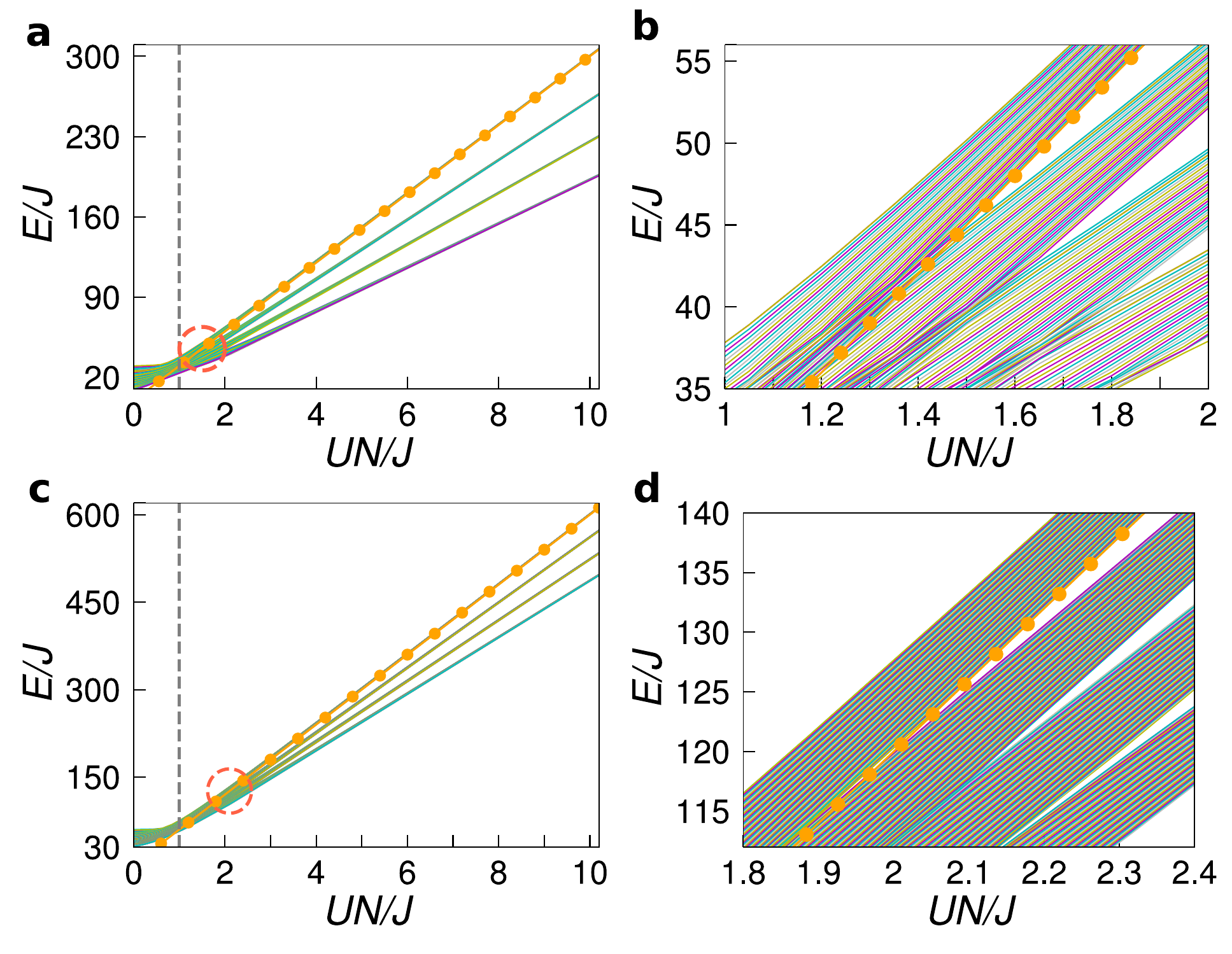}
 \caption{Energy level distributions. Results are shown for (\ref{hint}) in dimensionless units. Panels {\bf a}, {\bf b}: $N=30$. Panels { \bf c}, {\bf d}: $N=60$. The region marked with a circle on the panels {\bf a}, {\bf c} is enlarged in the panels {\bf b}, {\bf d}. The dashed vertical lines mark the threshold point, $UN/J=1$, separating Rabi and Josephson regimes. The ball lines mark the expectation energy $E=UN^2$ of $|N,0,0,\rangle$. Only the four highest energy bands are plotted.}
\label{level_spacing_2}
\end{figure}

Adopting the same classification for the integrable three-well system given by (\ref{hint}), numerical computation of the energy spectrum shows that transition from the Rabi to the Josephson regime is accompanied by the emergence of energy bands. Illustrative results are  depicted in Fig. \ref{level_spacing_2}. Hereafter units are chosen such that $\hbar=1$, and for all figures isotropic tunneling $J_1/J=J_3/J=1/\sqrt{2}$ is adopted for simplicity. 
The Hamiltonian acts on the Fock space spanned by the
normalized vectors  
$|{\mathcal N}_1,{\mathcal N}_2,{\mathcal N}_3 \rangle ={\mathcal C}^{-1} (a_1^\dagger)^{{\mathcal N}_1} (a_2^\dagger)^{{\mathcal N}_2} (a_3^\dagger)^{{\mathcal N}_3} |0\rangle, $
where ${\mathcal C}= \sqrt{{\mathcal N_1}!{\mathcal N_2}!{\mathcal N_3}!} $ and $|0\rangle \equiv |0,0,0\rangle$ is the Fock vacuum. 
On each panel   the line $E=UN^2$ is depicted. This quantity is the expectation value of the state $|N,0,0\rangle$. In the extreme Rabi regime with $U=0$ the energy levels are uniformly distributed with spacing $\Delta E = J$. The line $E=UN^2$ emerges from the midpoint of the entire energy spectrum when $U=0$, to lie on the lower edge of the uppermost energy band as $U$ is increased to bring the system into the Josephson regime. Note that the separation into distinct energy bands becomes very evident once the system is deep into the Josephson regime. These features significantly influence the dynamical evolution of the system from the initial state $|N,0,0\rangle$. In the Rabi regime, an accurate description of the initial state requires a linear combination over all eigenstates. 
However in the Josephson regime the state $|N,0,0\rangle$ can be accurately approximated as a linear combination of a subset of eigenstates, due to the band structure. This conclusion applies for all particles numbers, with the result represented in Fig. \ref{level_spacing_2} depicting the cases $N=30$ and $N=60$. Provided $UN/J \gg 1$, the separation into bands is clearly identifiable.  The consequences will be investigated at a deeper level in the next section, where we will fix $N=60$. Moreover, it will be shown how the breaking of the integrability, through the application of an external field, allows for control of the dynamics in a predictable fashion.     


\subsection*{Quantum dynamics}

The time evolution of the expectation values for the number operators are computed using 
$ \langle N_i\rangle = \langle\Psi(t)|N_i|\Psi(t)\rangle$,  $i=1,2,3$,
where $|\Psi (t)\rangle = \exp(-iHt)|\phi\rangle$ 
and $|\phi\rangle$ represents an initial state.  We adopt a protocol for which $|\phi\rangle =|N,0,0\rangle$, so the well labeled 1 is the source, the well labeled 2 is the gate, and the well labeled 3 is the drain.


\begin{figure}[b!]
\centering
\includegraphics[width=8cm]{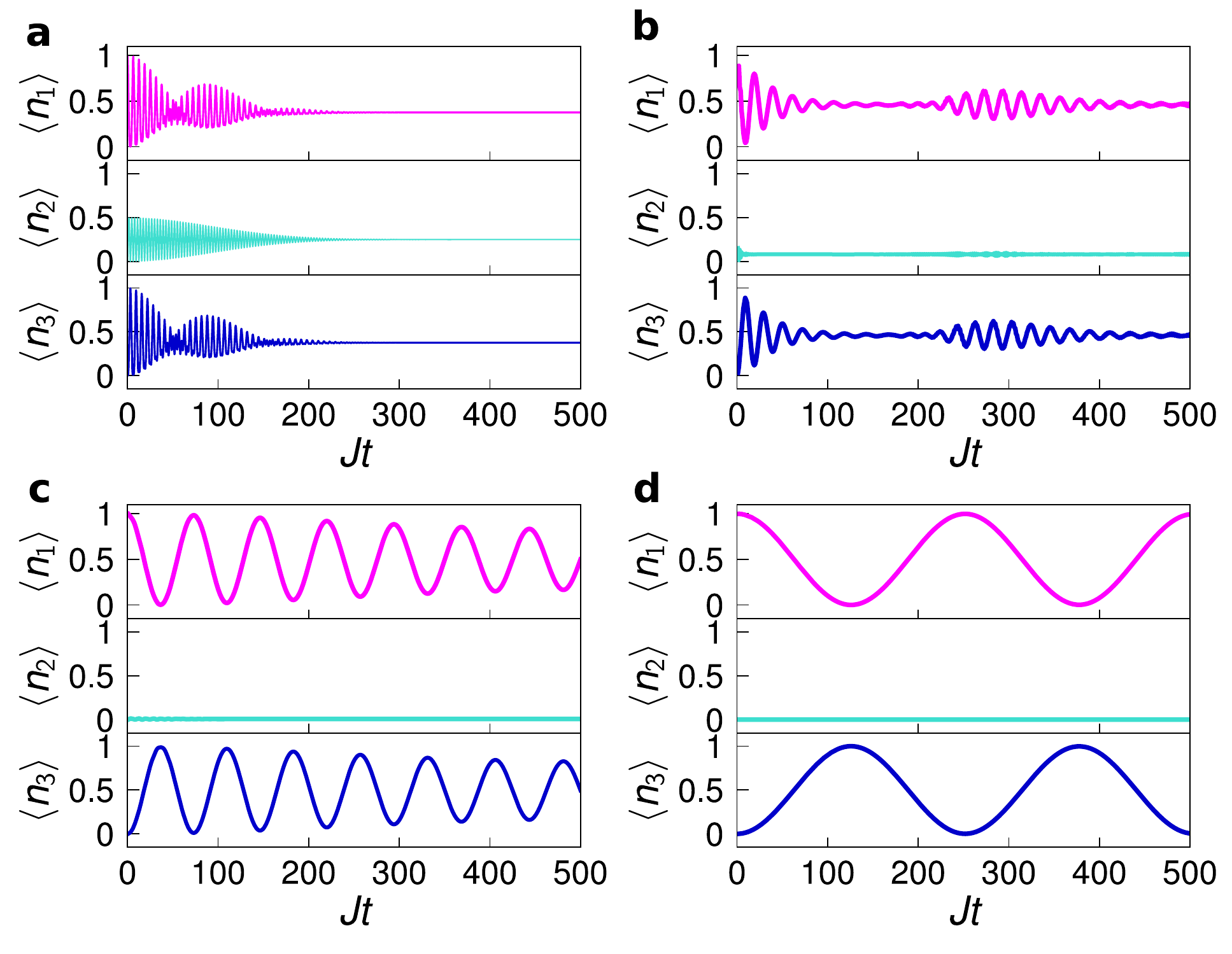}
      \caption{Time evolution of expectation values. Dimensionless units are used. In the integrable regime, the expectation values of the number operators are shown, with the initial state $|60,0,0\rangle$.  The configuration used has
$\epsilon/J =0$. {\bf a} $U/J=0.001$;
{\bf b} $U/J=0.015$; {\bf c} $U/J=0.05$; {\bf d} $U/J=0.17$.  It is apparent that increasing $U$ leads to an increasing suppression of tunneling into the gate, while maintaining oscillations between the source and the drain. In the case  {\bf d} the expectation value of the number operator associated with the gate is negligible, so tunneling to the gate is considered to be switched-off. The oscillations between the source and the drain are close to being harmonic and coherent.}
\label{din_int}
\end{figure}

\begin{figure}[t!]
\centering
\includegraphics[width=8cm]{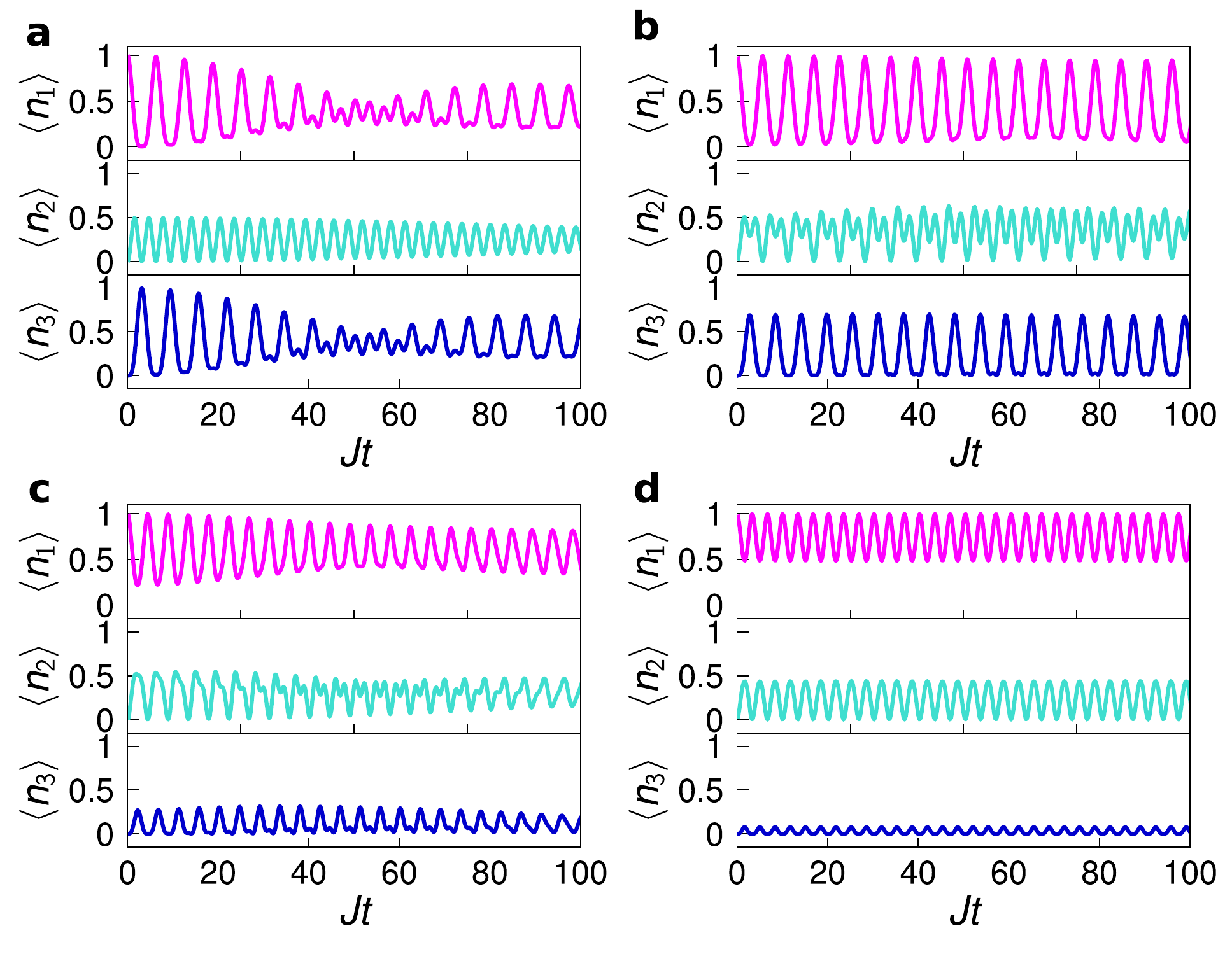}
\caption{Time evolution of expectation values. Dimensionless units are used. The effects of breaking integrability for the expectation values of the number operators are shown, with the initial state $|60,0,0\rangle$.  The configuration used has
$U/J=0.001$.   {\bf a} $\epsilon/J =0$;
 {\bf b} $\epsilon/J =0.47$;  {\bf c} $\epsilon/J =1.0$;  {\bf d} $\epsilon/J =1.63$. It is apparent that increasing $\epsilon $ leads to greater suppression of tunneling into the drain. For  {\bf d} the expectation value of the number operator associated with the drain is negligible, so tunneling into the drain is considered to be switched-off.} 
\label{transistor_2}
\end{figure}

We begin with  the integrable model (\ref{hint}) and first consider variations in the interaction parameter $U$ to manipulate the tunneling across the wells. Fig. \ref{din_int} presents results obtained for four choices of $U$. The dynamics typically display collapse and revival of oscillations in the Rabi regime, as in Fig. \ref{din_int}a.  
On increasing $U$, the period increases while the time-average of $\langle N_2 \rangle$ decreases.  Furthermore, the dynamics between wells 1 and 3 approach harmonic oscillations with $\langle N_1\rangle +\langle N_3\rangle \simeq N$. The transition between the Rabi and Josephson regimes can be seen, qualitatively, in the passage from Fig. \ref{din_int}b to Fig. \ref{din_int}c. This change in behaviour is in accord with the threshold point in Fig.\ref{level_spacing_2}.

In this latter regime, Fig. \ref{din_int}c and \ref{din_int}d,  these dynamical features can be understood by observing that the integrable Hamiltonian possesses a hidden two-well algebraic structure,
with an effective well given by the combined source and drain. As is well known \cite{milburn,l01,tlf05}, the self-trapping regime is expected to occur in the two-well model in the Josephson regime. 
To be more precise, $\langle N_2\rangle/N <\tilde{\epsilon}$ when $UN>J/(2\sqrt{\tilde{\epsilon}-\tilde{\epsilon}^2})$ if well 2 is initially empty. Thus, for $UN\gg J$ we find $\langle N_2\rangle/N \simeq 0$, and almost all bosons are distributed between the source and the drain if only a small fraction of bosons are initially in the gate.

On the other hand, 
it has been pointed out \cite{lps10} that for isotropic tunneling the source and the drain can form an effective non-interacting two-well system, by second-order processes \cite{Lukin, Kuklov,Bloch} through the gate, such that  $\langle N_2\rangle \simeq 0$. For general tunneling, we find the remarkable result that the effective Hamiltonian is simply given by $H_{\rm eff}=-\lambda Q$, where $Q$ is the conserved charge (\ref{charge}), and $\lambda^{-1}=4U(N-1)$ (details are provided in Methods). This produces an effective tunneling coupling given by $J_{\rm eff}=\lambda J_1J_3$, which decreases with increasing $N$, and therefore will only be observed in  mesoscopic samples \cite{lps10}.  In view of the above observations, 
we formally identify the resonant tunneling regime of the system to be determined by $UN \gg J $, which contains the Josephson regime.

In Fig. \ref{transistor_2} the time evolution of expectation values for number operators is displayed in a case of broken integrability. Increasing the value of $\epsilon $ suppresses the tunneling of particles into the drain, while increasing the time-average value of $\langle N_2 \rangle$. For $\epsilon/J =1.63$ this suppression of tunneling into the drain is strong enough that its number expectation value is close to negligible, i.e. tunneling into the drain has been switched-off. 


\subsection*{Control of resonant tunneling}

In Fig. \ref{din_int}d the dynamics is seen to be remarkably close to being harmonic over sufficiently  short time scales, with the period monotonically increasing with interaction coupling $U$. This behavior supports the conclusion that the effective Hamiltonian for the resonant tunneling regime is simply related to the conserved charge $Q$. The frequency of oscillation in this regime is given by $\omega_\mathrm{J} = \lambda J^2$, with the amplitude also being $U$-dependent. 
When the initial state is $|N,0,0\rangle$, the oscillations between the source and drain are coherent, 
with tunneling to the gate switched-off. On the other hand, if the initial state is $|0,N,0\rangle$ the system will remain trapped in this initial state configuration,
with tunneling from the gate switched-off.

Next, we maintain the system in the resonant tunneling regime $UN\gg J$ and study the non-integrable dynamics using the parameter $\epsilon $ to control the behavior of the source and drain subsystem. The approach here, following the study above, is to choose the initial state $|N,0,0\rangle$  and investigate the ability to control the frequency and amplitude of the populations oscillating between the source and the drain. 
\begin{figure}[h!]
\centering
\includegraphics[width=8cm]{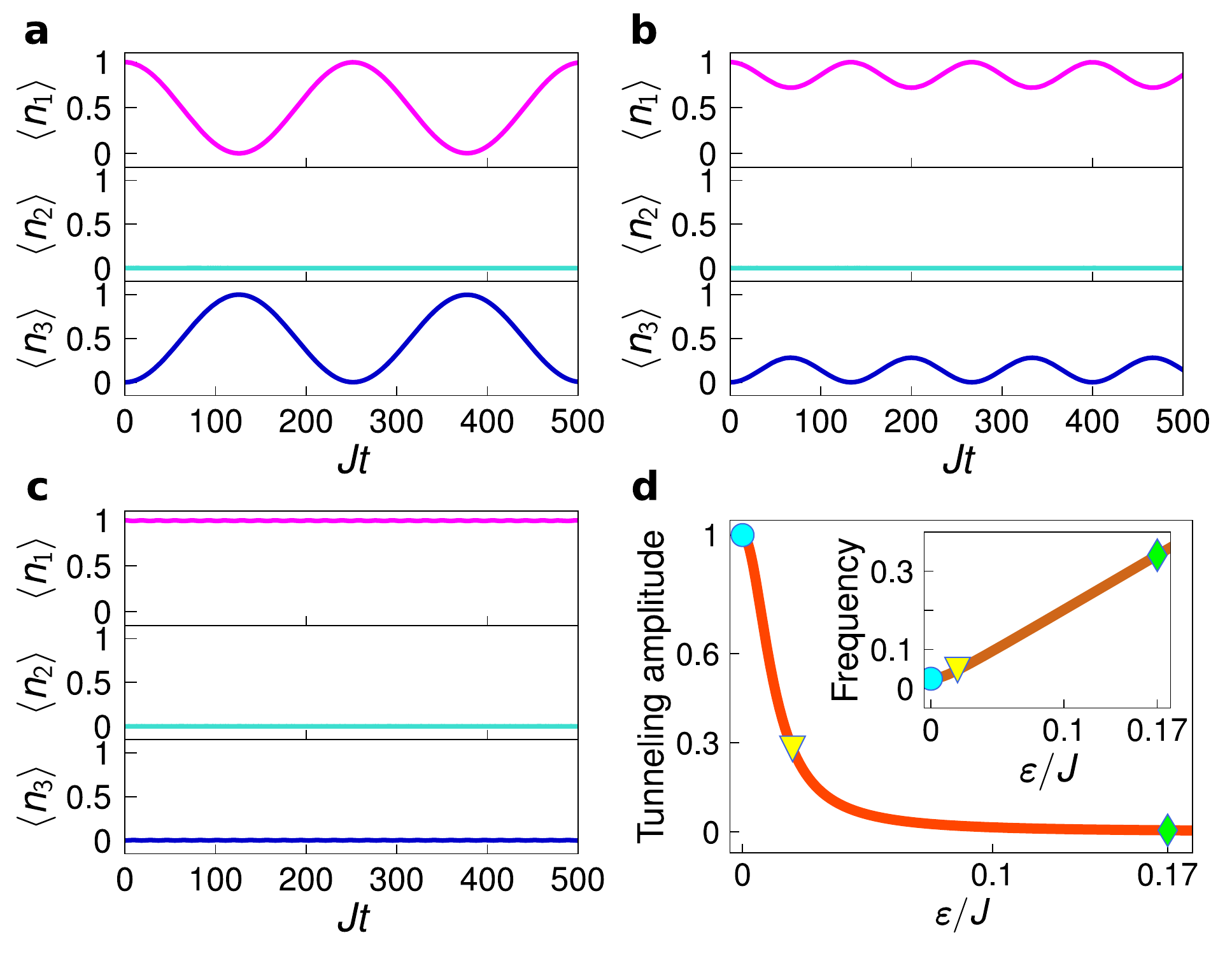}
\caption{Amplitude and frequency of oscillations. The configuration used has $N=60$, $U/J=0.17$, 
and initial state $|60,0,0\rangle$.  {\bf a} Number operator expectation values for $\epsilon/J =0$: switched-on configuration, with
maximum tunneling amplitude between the source and the drain.
 {\bf b} Expectation values for $\epsilon/J =0.02$: approximately 30\% of the maximal tunneling amplitude.  {\bf c} Expectation values for $\epsilon/J =0.17$: switched-off configuration.  {\bf d} Tunneling amplitude as a function of the external field $\epsilon/J$. 
In the inset: tunneling frequency versus the external field $\epsilon/J$. 
The markers in the curves correspond to the values of the amplitude of Fig. \ref{transistor}{\bf a} (cyan circle), \ref{transistor}{\bf b} (yellow triangle), and \ref{transistor}{\bf c} (lime diamond).}
\label{transistor}
\end{figure}

In Fig. \ref{transistor}a, \ref{transistor}b, \ref{transistor}c the interaction coupling is fixed as $U/J=0.17$, and results are shown for the expectation values of the populations using three choices for $\epsilon $. It is seen that the presence of the external field does not significantly influence the gate, in the sense that it does not affect the negligible average population $\langle N_2 \rangle$. Fig. \ref{transistor}d shows how the amplitude decays while increasing the external field, as well as the dependence of the frequency. The three points highlighted in the curves correspond to the values of the amplitude and frequency of Figs. \ref{transistor}a (cyan circle), \ref{transistor}b (yellow triangle), and \ref{transistor}c (lime diamond).

In this non-integrable regime the effective  Hamiltonian  is given by 
\begin{align}
H_{\rm eff}=-\lambda Q+\epsilon (N_3-N_1).
\label{heff}
\end{align}
For short time scales the dynamics exhibits Josephson-like oscillation \cite{weiss} with frequency 
\begin{align}
\omega_\mathrm{J} = \frac{2\lambda J_1J_3}{\sqrt{\Delta n}},
\label{freq}
\end{align}
where $\Delta n =1/(1+\gamma^2)$ is the amplitude and $\gamma = (\lambda (J_1^2-J_3^2)-2\epsilon )/2\lambda J_1 J_3$ (see Methods for details).
Increasing the external field reduces the oscillation amplitude $\Delta n$ and the period between the source and the drain, until the amplitude of oscillation is completely suppressed i.e. all tunneling is switched-off, demonstrating various levels of control, especially in the range $0<\epsilon <0.2$. Through semiclassical analysis, one can obtain analytic expressions for the expectation values of the relative populations, $n_i\equiv N_i/N$ ($i=1,3$), in the wells 1 and 3, given by $\langle n_1 \rangle = 1-\langle n_3 \rangle$ and $\langle n_3 \rangle = \Delta n \sin^2(\omega_\mathrm{J} t/2)$ (see details in Methods).
In agreement with Chuang et al. \cite{cold_transistor}, the maximum amplitude is obtained when the field is small.


\section*{Discussion}
We have analyzed a model for boson tunneling in a triple-well system. This was conducted in both integrable and non-integrable settings through variation of coupling parameters. The model draws an analogy with a transistor through identification of the wells as the source, gate, and drain.  Our primary objective was to investigate how this model could be implemented as an atomtronic switching device. 

In the integrable setting we identified the resonant tunneling regime between the source and  drain, for which expectation values of particle numbers in the gate are negligible. 
Moreover, it was found that a conserved operator of the integrable system acts as an effective Hamiltonian, which predicts coherent oscillations. This is in agreement with observations from numerical calculations. 

We then broke integrability through application of an external field to the source and the drain. It was shown in Fig. \ref{transistor_2} that the applied field to the system, in Rabi regime, was able to switch-off tunneling to the drain. On the other hand, in the resonant tunneling regime, the field did not destroy the harmonic nature of the oscillations, but did influence the amplitude and frequency.  Increasing the applied field allowed for tuning the system from the switched-on configuration through to switched-off (Fig. \ref{transistor}). Results from semiclassical analyses produced formulae for the amplitude and frequency, which proved to be remarkably accurate when compared to numerical calculations. This demonstrates the possibility to reliably control the harmonic dynamical behavior of the model in a particular regime. A surprising feature of this result is that the ability to  control the system in a predictable manner arises through the breaking of integrability. Our results open possibilities for multi-level logic applications and consequently new avenues in the design of atomtronic devices.

It is important, finally, to comment on the limitations of a three-mode Hamiltonian in the description of cold atom systems in a triple-well potential. Contributions from higher energy levels of the single-particle spectrum cannot be ignored under certain coupling regimes. For example, the presence of the external field will ultimately lead to level crossings as the field strength is increased. In the case of the analogous double-well system, estimates for when this may occur have been formulated\cite{dhc07}. 
We have undertaken checks to confirm that it is indeed possible to avoid these undesired scenarios, within an experimentally feasible scenario. 
(See Supplementary Note 1, Supplementary Figure 1 for further details). However it is also noteworthy that it is possible to include three-body, and higher, interaction terms as 
corrections\cite{dlns18} to compensate for when the three-mode approximation breaks down.    

\section*{Methods}

In this section we provide the details concerning algebraic structures behind the model, and complementary semiclassical approximations, which were used to derive analytic expressions characterizing quantum control in the resonant tunneling regime. These expressions were found to give close agreement with results obtained by exact numerical diagonalization (see Supplementary Note 2, Supplementary Figures 2 and 3 for more details).
We also discuss the feasibility of physical realization of the system.

\subsection*{Two-mode structure and the resonant tunneling regime}\label{algebra}
The integrable three-well model can be structured through two modes, as follows.
From Eq. \ref{hint}
we define $J=\sqrt{J_1^2+J_3^2}$ and  the operators $N_{1,3}=N_1+N_3$, $a_{1,3} = J^{-1}(J_1a_1+J_3a_3)$ and 
$a_{1,3}^\dagger = J^{-1}(J_1a_1^\dagger+J_3a_3^\dagger)$
satisfying the Heisenberg algebra
\begin{align*}
[N_{1,3}, a_{1,3}]=-a_{1,3}, & &[N_{1,3},a_{1,3}^\dagger]=a_{1,3}^\dagger, & &[a_{1,3}, a_{1,3}^\dagger]=1.
\end{align*}
Then
\begin{eqnarray} 
H_0 = U(N_{1,3}-N_2)^2 +J(a_{1,3}^\dagger a_2 +a_2^\dagger a_{1,3}),\nonumber
\end{eqnarray} 
such that the modes of wells 1 and 3 are now represented by the single mode ``$1,3$''.

The two-well model exhibits a self-trapping regime, with onset in the vicinity of $\chi \equiv UN/J\simeq 1$ \cite{milburn,tlf05}. This translates to a resonant tunneling regime for the triple-well model. Here we follow the approach of using semiclassical analysis\cite{rilf17}, such that  this regime may be clearly identified. Using the usual number-phase correspondence, that is,  
 $a_2=e^{i\theta_2}\sqrt{N_2}$, $a_{1,3}=e^{i\theta_{1,3}}\sqrt{N_{1,3}}$ and the conservation of boson number $N_{1,3}+N_2=N$, we find
\begin{eqnarray}
h=\frac{H_0}{N}=UN(1-2n_2)^2+2J\sqrt{(1-n_2)n_2}\cos\phi,\nonumber
\end{eqnarray}
where $n_2=N_2/N$ and $\phi = \theta_{1,3}-\theta_2$. Consider the dynamics where the initial condition is 
$n_2=0$. At the initial time $t=0$ the system has the energy $h=UN$. By energy conservation at all times,  we obtain the expression
\begin{eqnarray} 
n_2 = \frac{1}{2}-\frac{\sqrt{\chi^2-\cos^2\phi}}{2\chi}.
\label{n2}
\end{eqnarray}  
The conditions $\chi > 1$ and $|\cos\phi|=1$ (maximum value) imply that $0 \leq n_2 \leq 0.5$. From Eq. \ref{n2}, we conclude that when $\chi  \gg  |\cos\phi|$, $n_2\rightarrow 0$,
and the bosons are distributed between the wells labeled 1 and 3, producing the resonant tunneling regime. 

\subsection*{Effective integrable Hamiltonian for resonant tunneling}\label{heff_int}
In order to better understand the dynamics in the resonant tunneling regime, we first observe that the integrable Hamiltonian can be written, by using the conserved quantity $N$, 
as an effective Hamiltonian without on-site interaction (up to a global constant $UN^2$).  Specifically $H_0 = H_I+V$, 
where the interaction term $H_I=-4UN_2(N_1+N_3)$ has eigenstate and eigenvalues given by
\begin{eqnarray}
H_I|{\mathcal N}_1,{\mathcal N}_2,{\mathcal N}_3\rangle = -4U{\mathcal N}_2({\mathcal N}_1+{\mathcal N}_3)|{\mathcal N}_1,{\mathcal N}_2,{\mathcal N}_3\rangle, \nonumber
\end{eqnarray}
and the tunneling term $V=(J_1a_1^\dagger+J_3a_3^\dagger) a_2+{\rm h.c.}$ is treated as a perturbation.
For the isotropic case $J_1=J_3=J/\sqrt{2}$, \cite{lps10}, since $n_2\simeq 0$ the interaction 
 part is $H_I\simeq 0$ and the wells 1 and 3 form an effective non-interacting two-well system coupled through well $2$ by a 
 second-order process \cite{Lukin,Kuklov,Bloch,lps10} with the effective Hamiltonian 
$H_{\rm eff}=J_{\rm eff}(a_1^\dagger a_3+a_3^\dagger a_1)$.
Recall that the transition rate from initial state $|s\rangle$ to final state $|k\rangle$ 
is expressed 
\begin{align*}
W^{(i)}={2\pi}|\langle k|V^{(i)}|s\rangle|^2\delta(E_k-E_s), \qquad i=1,2,
\end{align*}
where $V^{(1)}=V$ for first-order transition (Fermi's golden rule), $\delta$ is the delta function and 
\begin{align*}
V^{(2)} =\sum_m\frac{V|m\rangle\langle m|V}{E_s-E_m}
\end{align*}
for second-order transitions. 
Equating second-order transition of $V$  with the
first-order transition of $H_{\rm eff}$ for the states $|N,0,0\rangle$ and $|N-1,0,1\rangle$, it is found that 
\begin{align*}
J_{\rm eff}=\frac{J^2}{8U(N-1)}.\nonumber
\end{align*}
Observing that $J_3^2 N_1+J_1^2N_3$ is constant for isotropic tunneling in the regime $\chi \gg 1$, then it does not affect the dynamics if we consider the linear combination $J_{\rm eff}(a_1^\dagger a_3+a_3^\dagger a_1)+\lambda' (J_3^2 N_1+J_1^2N_3)$. By numerical inspection we conclude that the effective Hamiltonian for general tunneling, which includes the anisotropic tunneling $J_1\neq J_3$, is given by
\begin{align*}
H_{\rm eff} =-\lambda Q,
\end{align*}
where $Q=J_1^2N_3+J_3^2N_1-J_1J_3(a_1 a_3^\dagger+a_1^\dagger a_3)$ is conserved and $\lambda^{-1} = {4U(N-1)}$. 

\subsection*{Effective non-integrable Hamiltonian and quantum control}\label{heff_nonint}
For the non-integrable case, the effective Hamiltonian in the resonant tunneling regime $\chi \gg 1$ is given 
by $H_{\rm eff}=-\lambda  Q+\epsilon( N_3-N_1)$. Returning to a semiclassical analysis it is found that, up to an irrelevant constant, 
\begin{align}
h& =\frac{H_{\rm eff}}{N} = - \lambda J_1^2(1-n_1)-(\lambda J_3^2+2\epsilon)n_1 \nonumber \\ 
&\hspace{2cm}+ 2\lambda J_1J_3\sqrt{n_1(1-n_1)}\cos\varphi,\nonumber
\end{align}
where $n_1=N_1/N$ and $\varphi = \theta_1-\theta_3$. For initial condition $n_1=1$ and $n_3=0$ we
have $h=-\lambda  J_3^2$, a constant. Applying  energy conservation and the condition $\cos\varphi=\pm 1$, we find that the amplitude of oscillation $\Delta n$ (Fig. 5d) is given by
$\Delta n=1/(1+\gamma^2)$,
where $\gamma =[\lambda(J_1^2 -J_3^2)-2\epsilon]/2\lambda J_1 J_3$.
Hamilton's equation gives
\begin{align*}
\dot{N}_1 =-\frac{\partial H_{\rm eff}}{\partial \theta_1}&=2\lambda J_1J_3\sqrt{N_1(N-N_1)}\sin\varphi, \\
\Rightarrow \qquad  \dot{n}_1 &= 2\lambda J_1J_3\sqrt{n_1(1-n_1)}\sin\varphi.
\end{align*}
Using the {\it Ansatz} $n_1=1-\Delta n \sin^2(\eta t)$,
we can easily verify that the above results provide analytic expressions for the expectation values   
\begin{eqnarray*}
\langle n_1\rangle = 1-\Delta n \sin^2\left(\omega_\mathrm{J} t/2\right),\quad
\langle n_3\rangle = \Delta n \sin^2\left(\omega_\mathrm{J} t/2\right),
\label{n1_n3}
\end{eqnarray*}
where $\omega_\mathrm{J}$ is the frequency given by Eq. \ref{freq}. 
Results for similar types of investigation have been presented in the case of pair-tunneling between two wells\cite{Bloch}.

\subsection*{Physical realization}

Here we discuss the feasibility of physical realization of the 
triple-well Hamiltonian (\ref{hint}), through use of Bose-Einstein condensates (BECs) of dipolar atoms.

\begin{figure}[h!]
\centering
\includegraphics[width=8.3cm]{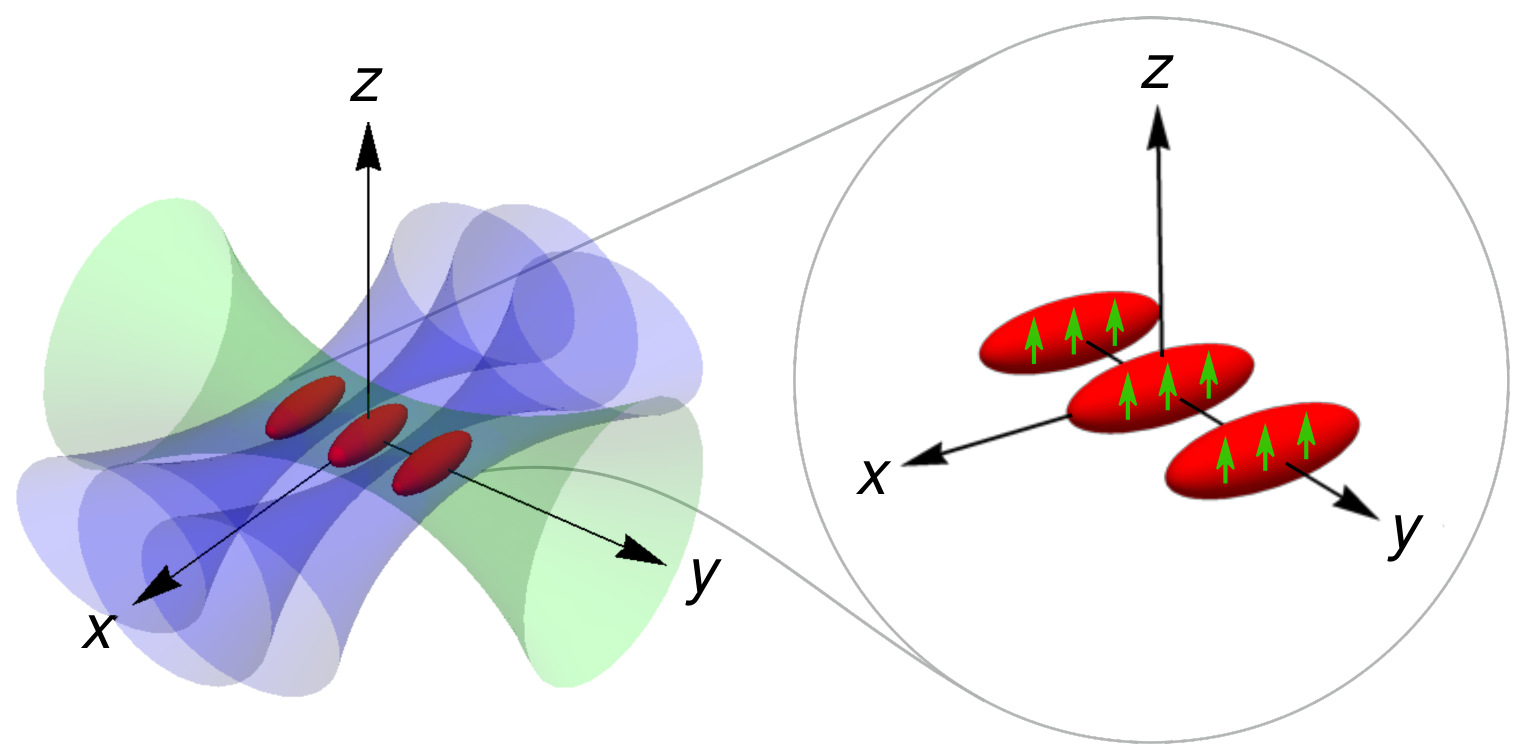}
\caption{Schematic representation of the trap geometry. Three parallel lasers (blue) are crossed by a transverse beam (green). The cigar-shapes, in red, represent a dipolar Bose-Einstein condensate trapped in a triple-well potential, and the green internal arrows depict the orientation of the dipoles. 
}
\label{bec_esq}
\end{figure}

Three parallel, tightly focused Gaussian beams, with waist of 1 $\mu$m and wavelength $\lambda = 1.064$ nm, separated by a distance $l=1.8\,\mu$m, form an optical triple-well potential aligned along the 
$y$-axis \cite{lps10}. A transverse beam, with waist of 6 $\mu$m, provides  $xz$-confinement. For such a setup, in the harmonic approximation the potential of each well $i=1,2,3$ is symmetrically cylindrical and is given by
\begin{eqnarray*}
V_{0}(x,y,z) = \sum_{i=1}^3\left(\frac{1}{2}m\omega_x^2 x^2+\frac{1}{2}m\omega_r^2((y-y_i)^2+z^2)
\right),
\end{eqnarray*}      
where $y_i=l,0,-l$. The trap frequencies $\omega_x$ and $\omega_r$ can be controlled by the intensity of the laser beams. 
This configuration facilitates the formation of three cigar-shaped wells.
We consider a system of bosons with dipole-dipole interactions to provide long-range interactions, and weakly repulsive contact interactions to promote condensate stability \cite{Griesmaier}. The dipoles are  oriented along the $z$-direction. See Fig. \ref{bec_esq}. 

The transverse beam performs the function of the external field that controls the device. Its focus, when displaced along the $y$-axis by $\Delta y$, introduces  the potential energy
\begin{eqnarray*}
V_1(y) =  \frac{1}{2}m\omega_y^2\Delta y(2y+\Delta y).
\end{eqnarray*}    
This generates a potential difference, resulting in the external field strength $\epsilon = m\omega_y^2\Delta y  l$. 
The frequency of the transverse laser ($\omega_y$) is much lower than the frequency of the parallel lasers ($\omega_r$), so that displacement by $\Delta y$ 
introduces a ``tilting'' of wells 1 and 3. 
These are the relevant wells in the resonant  tunneling regime.

For the case of a dipolar BEC of $^{52}$Cr \cite{LS_2009}, we numerically find that the integrability condition, with $\alpha \sim 5.8$, is achieved for $\omega_x \sim 2 \pi \times 64$ Hz, $\omega_r \sim 2\pi \times 220$ Hz, where we assumed the Gaussian approximation for the ground state. The value of $U$ obtained in units of $J$ is $U/J \sim 7.5\times 10^{-3}$, which means that the resonant tunneling regime can be achieved for $N\gg 130$ atoms. In principle, this condition can be satisfied experimentally \cite{LS_2009}. 
As an example, for $N\sim 5000$ atoms, with $\omega_y \sim 2\pi \times 1.5$ Hz, translating the transverse laser by $\Delta y = 1.8\,\mu$m we obtain $\epsilon/h \sim 3.6\times 10^{-2}$ Hz. It results in a tunneling amplitude $\Delta n \sim 0.25$, which means that $25\%$ of the population of the source in the initial state $|N,0,0\rangle$ switch to the drain, and back, through harmonic tunneling. This example approaches the case of Fig. \ref{transistor}b.

Another strongly dipolar BEC which can be considered is  $^{164}$Dy \cite{barbut16}. In this case, we calculate $\alpha \sim 5.9$ for $ \omega_x \sim 2 \pi \times 22.7$ Hz, $\omega_r \sim 2\pi \times 67.3$ Hz with $U/J \sim 2.2\times 10^{-2}$. For $N\sim 500$ atoms, with $\omega_y \sim 2\pi \times 0.76$ Hz, this yields $\epsilon/h \sim 3.0\times 10^{-2}$ Hz and $\Delta n \sim 0.23$. 

In both cases above, the parameter choices are such that higher-order interaction terms, such as correlated hopping, are negligible \cite{mgi06}.

An analysis of the effects of perturbations is provided in Supplementary Note 3: Fidelity dynamics, Supplementary Figures 4 and 5.

\subsection*{Competing interests:} The authors declare no competing interests.

\subsection*{Data availability:} All relevant data are available on reasonable request from the
authors.

\subsection*{Author contributions:} All authors contributed to the conceptualisation of the project, and actively engaged in the writing of the manuscript.
K.W.W., L.H.Y., and A.P.T.  implemented the theoretical analyses of the model,  detailed the experimental feasibility, and  processed the numerical computations.
J.L. and A.F. designed the research framework, and directed the program of activities.


\section*{Acknowledgements}
K.W.W. and A.F. were supported
by CNPq (Conselho Nacional de Desenvolvimento Cient\'ifico e Tecnol\'ogico), Brazil. J.L. and A.F. were supported by the Australian Research Council through Discovery Project DP150101294.  We thank Henri Boudinov, Ricardo R. B. Correia, Matt Davis, and Artem Volosniev for helpful discussions.

\newpage
\subsection*{Supplementary Note 1: Energy levels for the non-integrable Hamiltonian}

In Supplementary Figure 1, we show the behavior of the energy levels for the non-integrable Hamiltonian as we increase the external field $\epsilon$. This will allow us to determine the range of $\epsilon$ for which the switching device can operate.

In order to control the device without unwanted virtual processes, we need the parameter $\epsilon$ to be less than
a threshold value $\epsilon_\mathrm{\mathsf{c}}$, defined by the point where the crossing between the energy level bands starts. For the case where $|4U(N-1)\epsilon| \gg |J_1J_3|$, the crossing between the two upper bands occurs when
\begin{align*}
\Delta E &\simeq (E_0 - N\epsilon)-(E_1+(N-1)\epsilon) = 0 \nonumber \\
\Rightarrow \qquad \epsilon_\mathrm{\mathsf{c}} &\simeq 2U
\end{align*}
as can be seen in Supplementary Figure 1 by the brown dot-dashed lines.
It is worth noting that the gap between adjacent energy levels of the same band, $2 J_{\rm eff} \simeq 2 \epsilon$, increases with $\epsilon$. This increase causes a tendency toward self-trapping, as can be seen in the amplitude equation, $\Delta n=1/(1+\gamma^2)\simeq 1/(1+ (4U(N-1)\epsilon)^2/(J_1 J_3)^2)$, presented in Methods.

Considering switched-off states to be defined by the condition  $\Delta n < 1/N$, the $\epsilon$ dependency for the switched-off configuration becomes
\begin{align}
\frac{J_1 J_3}{4U\sqrt[]{(N-1)}} < \epsilon < 2U.
\label{epsilon_limits}
\end{align}
This means that the device's controlled tunneling occurs for $\epsilon < J_1 J_3/4U\sqrt[]{(N-1)}$, while the off-state range occurs between $J_1 J_3/4U\sqrt[]{(N-1)} < \epsilon < 2U$. For the case of Supplementary Figure 1b, the off-state range is given by $0.1 \leq \epsilon < 0.34$,  in accordance with Fig. 5 of the main article.
\begin{figure}[h!]
\centering
      \includegraphics[width=8.4cm]{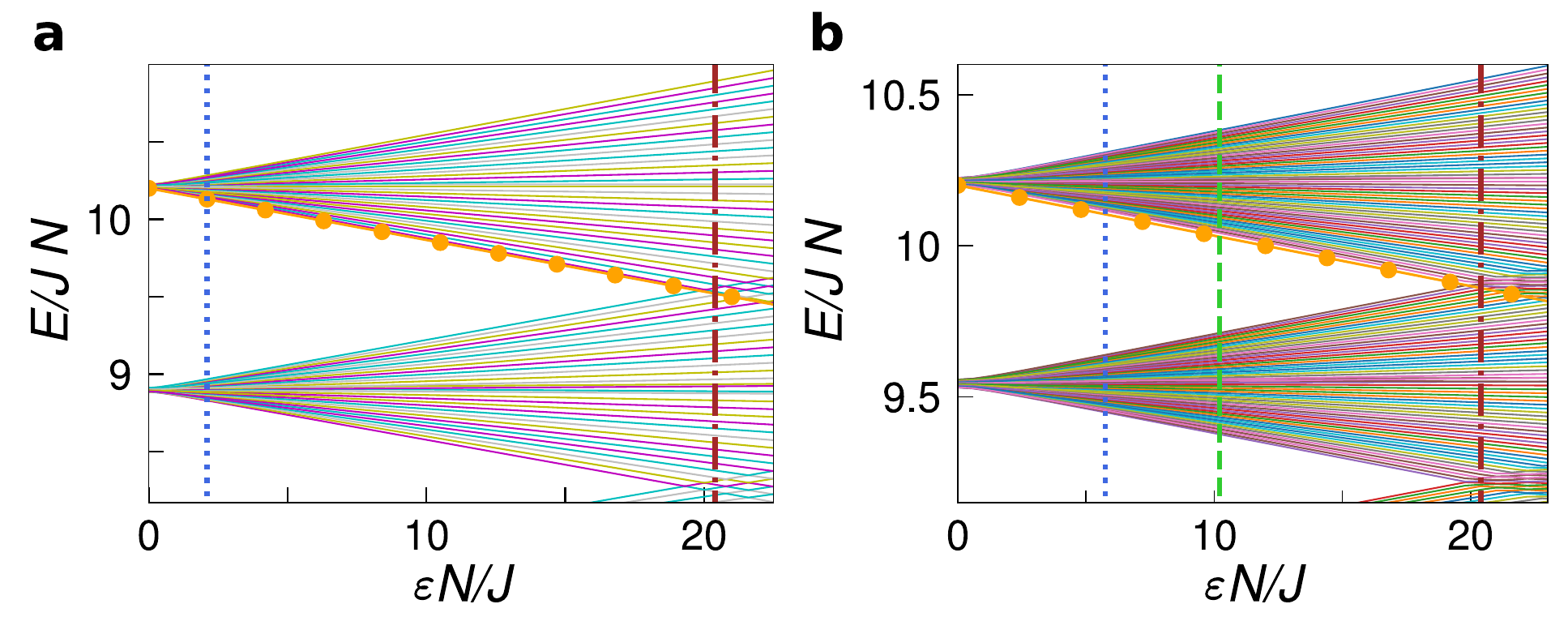}
      \caption*{
\textbf{Supplementary Figure 1.} Energy levels. $E/JN$ vs. external field $\epsilon N/J$. The orange ball lines mark the initial energy: $E=UN^2-N\epsilon$. According to Eq. \ref{epsilon_limits}, the blue dot line marks the region of the parameter $\epsilon$ for which the switched-off configuration starts, while the brown dot-dashed lines mark the threshold parameter $\epsilon_\mathrm{\mathsf{c}} \simeq 2U$, above which unwanted virtual tunneling may occur. The green dashed line corresponds to the switched-off point shown in Supplementary Figure 2 with the diamond mark.
The switching device operates between $0 \leq \epsilon < \epsilon_\mathrm{\mathsf{c}}$. {\bf a}: $N=30$ and $U=0.34$, $UN/J \simeq 10$. {\bf b}: $N=60$ and $U=0.17$, $UN/J \simeq 10$. Only the two highest energy bands are plotted.}
\label{fig_epsilon_c}
\end{figure}

\subsection*{Supplementary Note 2: Comparison between analytic expressions and numerical diagonalization}
Supplementary Figure 2 presents the period dependence on the external field obtained from the analytic expression for the frequency, Eq. 6 in the main article.  It compares  three points obtained from numerical analysis, giving support that the analytic expressions are in accordance with quantum dynamics. In Supplementary Figure 3, it is shown how the temporal evolution of amplitude oscillation  depends on the external field for the resonant tunneling case.

\begin{figure}[h!]
\centering
\includegraphics[width=5cm]{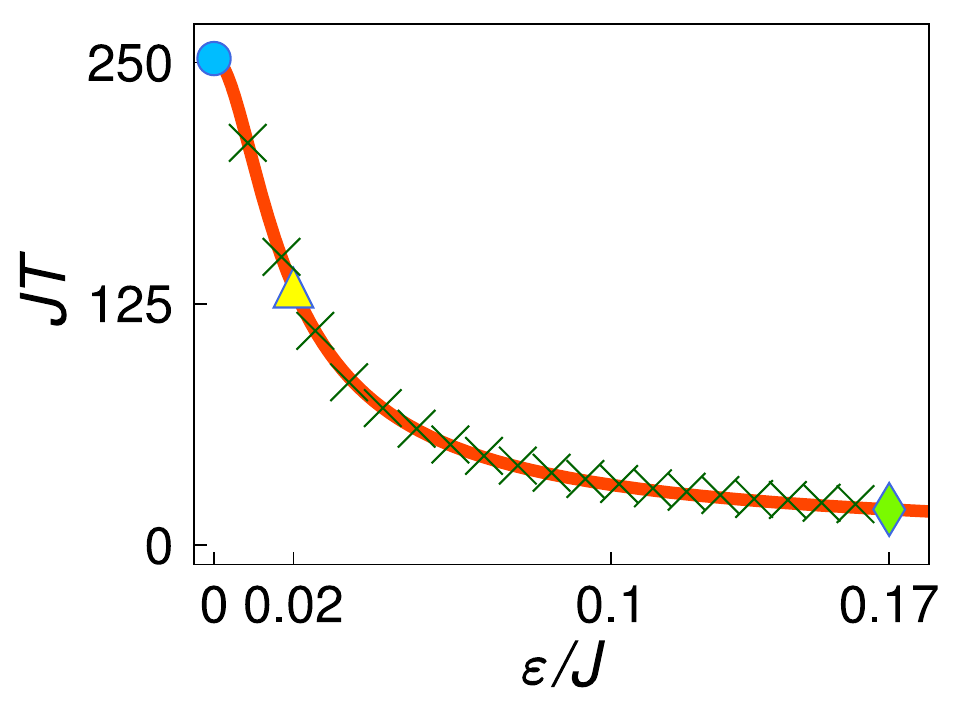}
\caption*{\textbf{Supplementary Figure 2.} Period of oscillation. $JT$ vs. the external field $\epsilon/J$. The orange line shows the period function obtained from the analytic expression through Eq. 6 in the main text, compared with numerical points marked in ``x''. The other markers correspond to the values of the period of Figs. 5a (blue circle), 5b (yellow triangle) and 5c (lime diamond) in the main article, also obtained from numerical analysis. The configuration used here has $N=60, U/J=0.17$, and initial state $|60,0,0\rangle$.}
\label{smfig}
\includegraphics[width=7cm]{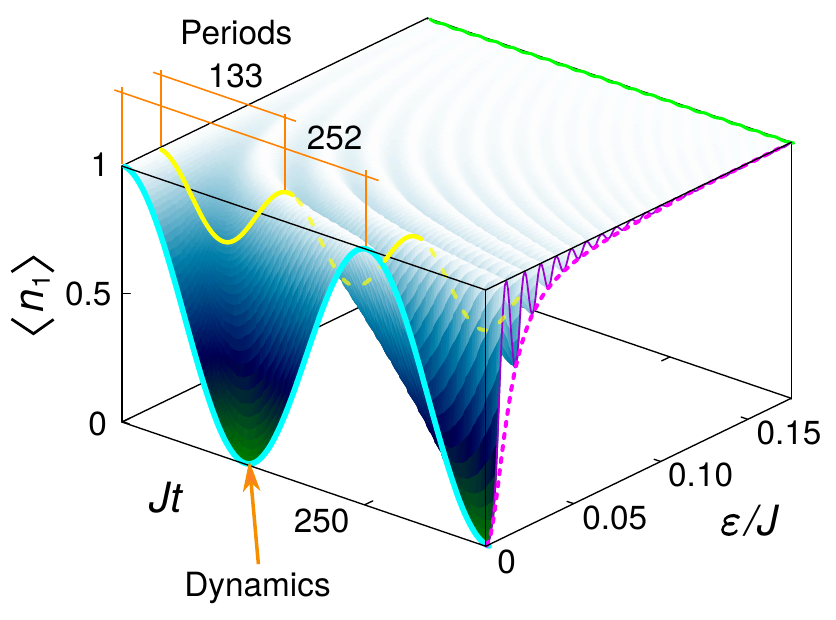}
\caption*{\textbf{Supplementary Figure 3.} Source dynamics. $\langle n_1\rangle$ vs. the external field $\epsilon/J$. The graph shows the harmonic dynamic evolution of the fractional occupation in well 1 (source) as a function of time and external field parameter $\epsilon$. The dynamics of $\epsilon/J = 0$ (cyan line), $\epsilon/J = 0.02$ (yellow line) and $\epsilon/J = 0.17$ (lime line) are highlighted and the dashed pink line is marking the amplitude function $\Delta n$.
The configuration used here has $N=60$, $U/J=0.17$, and initial state $|60,0,0\rangle$, as in  Fig. 5 of the main text. 
Recall that the dynamics of well 3 (drain) has the same configuration, however it is out of phase by $\pi$, while $\langle n_2 \rangle=0$.}
\label{3dfig}
\end{figure}

\newpage

\subsection*{Supplementary Note 3: Fidelity dynamics}
In using properties of integrability to draw conclusions about this model, it is important to understand the effects of perturbations away from the integrable point. It has been found that breaking the integrability, through an external field, still allowed for predictions of the amplitude and frequency in the resonant tunneling case. Here an extended analysis is presented to highlight the effects of another integrability-breaking perturbation.  

We use fidelity, which is computed through $F(t)=|\langle \psi_0 |\psi(t) \rangle |^2$ where $|\psi_0 \rangle=|N,0,0\rangle$, and time-evolution is governed by the Hamiltonian
\begin{align}
H = H_0 + \xi N_1 N_3 + \epsilon(N_3-N_1)
\label{hls_erro}
\end{align}
where $\xi$ is the perturbation parameter.
This quantity allows for an investigation of the probability to revisit the initial state configuration after some period of time evolution, complementing the earlier dynamics studies.

Supplementary Figure 4 presents fidelity dynamics in the integrable case for different values of the interacting parameters $U$. Clearly, the probability to find the initial state becomes relevant in the Josephson and Fock regimes, cf. Fig. 3 in the main article. In the resonant regime, the initial state can be recovered to a high degree of accuracy. As $\langle N_2\rangle \simeq 0$ in this regime, almost the entire boson population periodically tunnels back and forth.  The probability of the initial state $|N,0,0\rangle$ to be transferred to the drain as $|0,0,N\rangle$, is close to 1.

Supplementary Figure 5 illustrates the impacts of perturbation,
through $\xi \neq 0$, for the cases analogous to those of Fig. 5a, 5b in the main article. It is verified that increasing the value of the perturbation leads  to a decrease in the fidelity until there are no oscillations. This limit occurs when the perturbation is less than the order of $J_{\rm eff}$. Interestingly, when the 
external field $\epsilon$ is included (dotted lines of the graphs), it improves the fidelity  compared $\epsilon=0$ (solid lines of the graphs). Thus the external field term renders the harmonic dynamics of the system in the resonant tunneling regime to be more resistant to perturbation. 

In summary, the integrability breaking coupling $\epsilon$ does not destroy the harmonic nature of oscillations in the resonant tunneling regime. It does influence the period and amplitude, resulting in localization in wells 1 and 3. But this occurs in a fashion which is predictable, through the results of the classical analysis. In contrast,  the coupling $\xi$ does destroy the harmonic nature and produces more pronounced localization, however the effects are mitigated through increasing $\epsilon$. When the coupling $\xi$ is seen as a perturbation in the sense that $\xi \ll \epsilon$, the device dynamics are robust against this perturbation.

\newpage
\begin{figure}[h!]
\centering
\includegraphics[width=6.1cm]{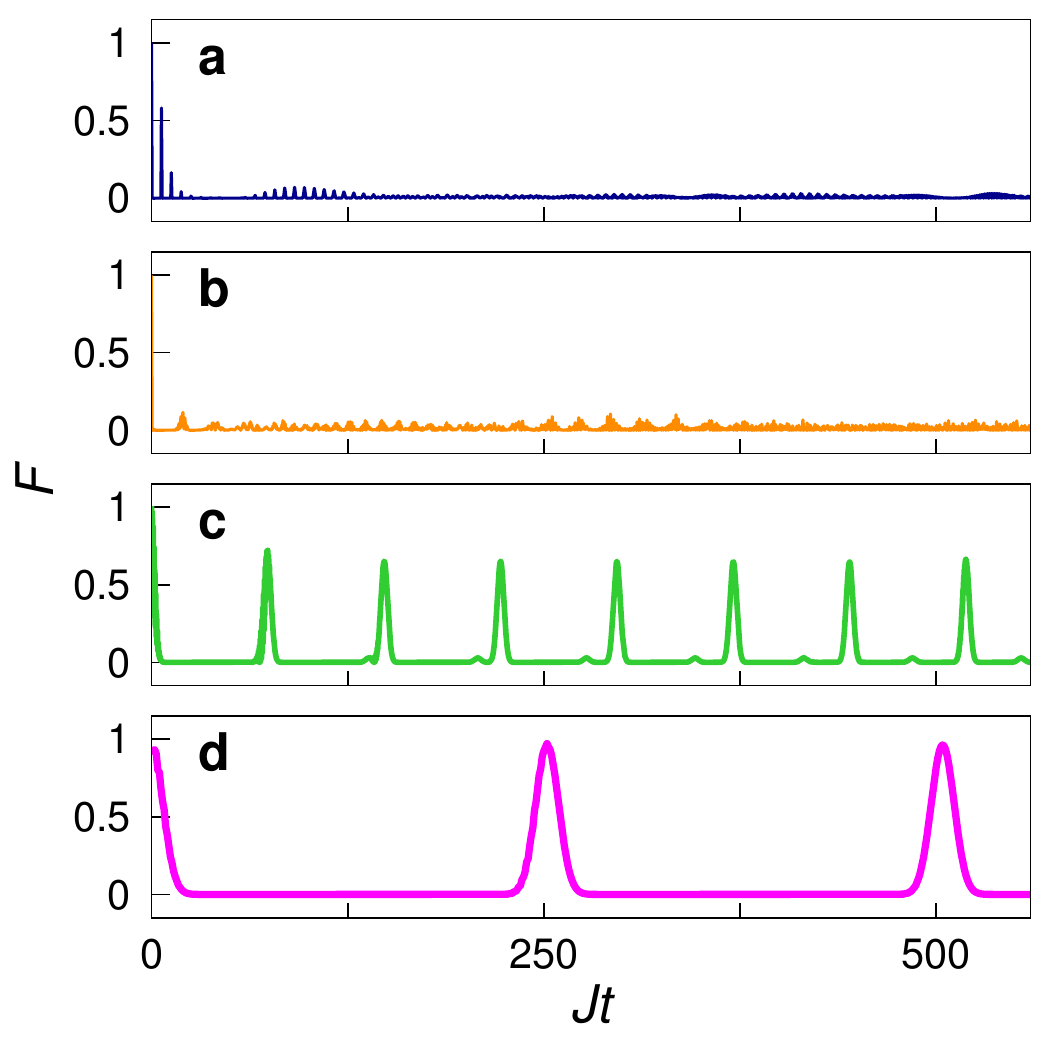}
\caption*{\textbf{Supplementary Figure 4.} Fidelity. $F$ vs. time $Jt$ for the Hamiltonian  Eq. \ref{hls_erro}, with $\xi/J=0$ and $\epsilon/J=0$, and four different parameter values of $U/J$ as in Fig. 3 in the main article. {\bf a} $U/J=0.001$, {\bf b} $U/J=0.015$, {\bf c} $U/J=0.05$, and {\bf d} $U/J=0.17$. 
The configuration has initial state $|60,0,0\rangle$. Observe the transition from the Rabi regime to Josephson from panel {\bf b} to panel {\bf c}.}
\label{fid_1}
\end{figure}

\begin{figure}[h!]
\centering
\includegraphics[width=6.1cm]{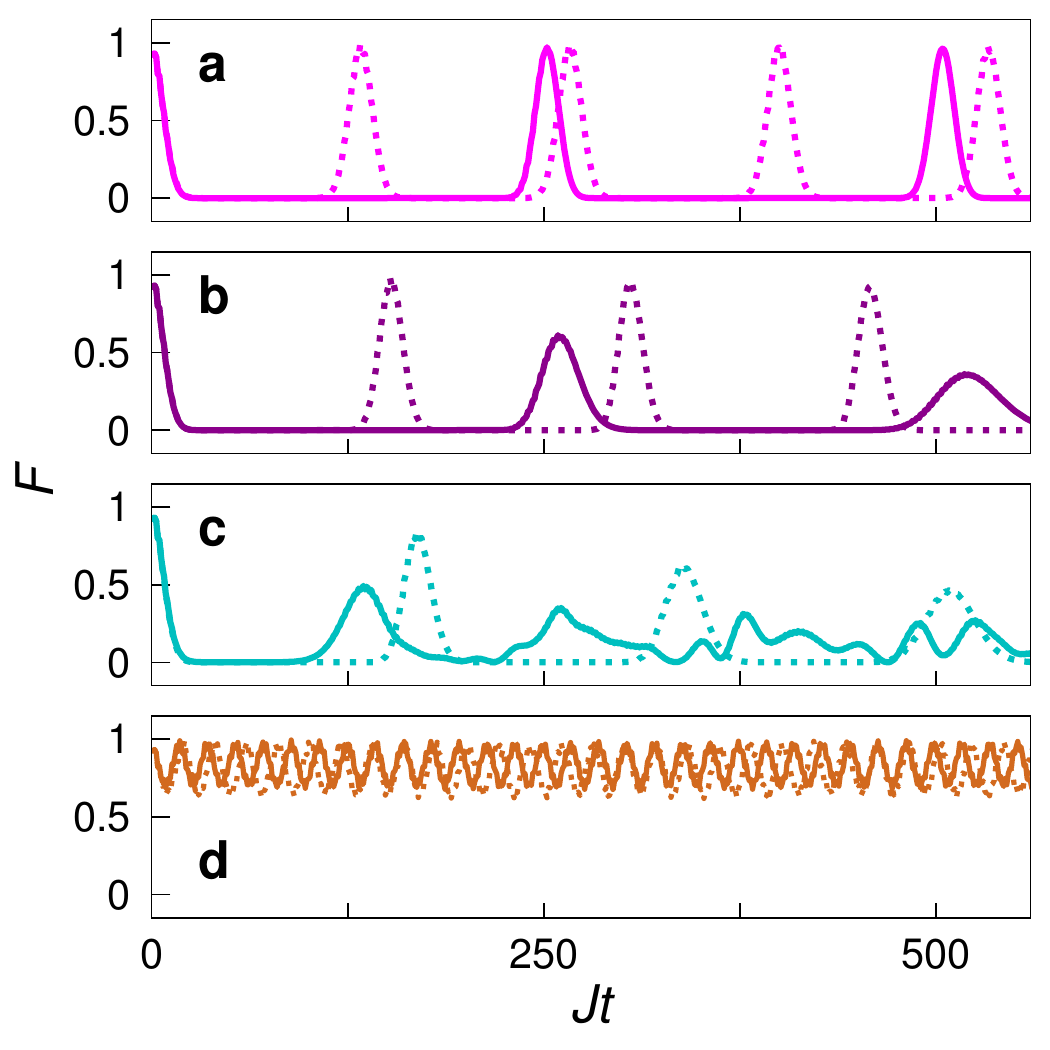}
\caption*{\textbf{Supplementary Figure 5.} Fidelity. $F$ vs. time $Jt$ for the Hamiltonian Eq.  \ref{hls_erro}, with $\epsilon/h=0$ (continuous line), $\epsilon/J=0.02$ (dotted line) and $\xi/J$ varying between panels. 
{\bf a} $\xi/J=0$, {\bf b} $\xi/J=0.0003$, {\bf c} $\xi/J=0.001$, {\bf d} $\xi/J=0.01 \simeq J_{\rm eff}/J$.
The configuration has initial state $|60,0,0\rangle$ and $U/J=0.17$. The parameters $\epsilon$ are chosen according to the values used in Fig. 5a and 5b in the main article.}
\label{fid_1_epsilon}
\end{figure}


\end{document}